\documentclass[journal=jacsat,manuscript=article]{achemso}
\usepackage[version=3]{mhchem} % Formula subscripts using \ce{}

\usepackage{amssymb}
\usepackage[dvipsnames]{xcolor}
\usepackage{comment}
\usepackage{soul}
\usepackage{lineno}
\usepackage{natbib}
\usepackage{hyperref}
\hypersetup{
  colorlinks=true,
  citecolor=blue,
  linkcolor=blue,  
  urlcolor=blue,
  }

\newcommand*{\citen}[1]{%
 \begingroup
  \romannumeral-`\x % remove space at the beginning of \setcitestyle
  \setcitestyle{numbers}%
  \cite{#1}%
 \endgroup  
}

\makeatletter
\newcommand{\undersim}[1]{\mathrel{\mathpalette\@undersim{#1}}}
\newcommand{\@undersim}[2]{%
 \vcenter{%
  \ialign{%
   ##\cr
   $\m@th#1#2$\cr
   \noalign{\nointerlineskip\kern.2ex}
   $\m@th#1\sim$\cr
   \noalign{\kern-.4ex}
  }%
 }%
}

\makeatother

\author{Cristina Sanz-Fern\'andez}
\affiliation{Centro de F\'{i}sica de Materiales (CFM-MPC), Centro Mixto CSIC-UPV/EHU, 20018 Donostia-San Sebasti\'{a}n, Basque Country, Spain} 
\altaffiliation{Contributed equally to this work}
\email{cristina_sanz001@ehu.eus}

\author{Van Tuong Pham}
\affiliation{CIC nanoGUNE, 20018 Donostia-San Sebasti\'{a}n, Basque Country, Spain}
\altaffiliation{Contributed equally to this work}
\email{v.pham@nanogune.eu}

\author{Edurne Sagasta}
\affiliation{CIC nanoGUNE, 20018 Donostia-San Sebasti\'{a}n, Basque Country, Spain}

\author{Luis E. Hueso}
\affiliation{CIC nanoGUNE, 20018 Donostia-San Sebasti\'{a}n, Basque Country, Spain}
\affiliation{IKERBASQUE, Basque Foundation for Science, 48013 Bilbao, Basque Country, Spain}

\author{Ilya V. Tokatly}
\affiliation{Nano-Bio Spectroscopy Group, Departamento de F\'isica de Materiales, Universidad del Pa\'is Vasco (UPV/EHU), 20018 Donostia-San Sebasti\'{a}n, Basque Country, Spain} 
\affiliation{IKERBASQUE, Basque Foundation for Science, 48013 Bilbao, Basque Country, Spain}
\affiliation{Donostia International Physics Center (DIPC), 20018 Donostia-San Sebasti\'{a}n, Basque Country, Spain}
\email{ilya.tokatly@ehu.es}

\author{F\`elix Casanova}
\affiliation{CIC nanoGUNE, 20018 Donostia-San Sebasti\'{a}n, Basque Country, Spain}
\affiliation{IKERBASQUE, Basque Foundation for Science, 48013 Bilbao, Basque Country, Spain}
\email{f.casanova@nanogune.eu}

\author{F. Sebasti\'an Bergeret}
\affiliation{Centro de F\'{i}sica de Materiales (CFM-MPC), Centro Mixto CSIC-UPV/EHU, 20018 Donostia-San Sebasti\'{a}n, Basque Country, Spain}
\affiliation{Donostia International Physics Center (DIPC), 20018 Donostia-San Sebasti\'{a}n, Basque Country, Spain}
\email{fs.bergeret@csic.es}
\title{
Quantification of interfacial spin-charge conversion in metal/insulator hybrid structures by generalized boundary conditions}

\abbreviations{IR,NMR,UV}

\begin{document}

\begin{abstract} 
We present and verify experimentally a universal theoretical framework for the description of spin-charge interconversion in non-magnetic metal/insulator structures with interfacial spin-orbit coupling (ISOC). Our formulation is based on drift-diffusion equations supplemented with generalized boundary conditions. The latter encode the effects of ISOC and relate the electronic transport in such systems to spin loss and spin-charge interconversion at the interface, which are parameterized, respectively, by $G_{\parallel/\perp}$ and $\sigma_{\rm{sc/cs}}$. We demonstrate that the conversion efficiency depends solely on these interfacial parameters. 
We apply our formalism to two typical spintronic devices that exploit ISOC: a lateral spin valve and a multilayer Hall bar, for which we calculate the non-local resistance and the spin Hall magnetoresistance, respectively. Finally, we perform measurements on these two devices with a BiO$_x$/Cu interface and verify that transport properties related to the ISOC are quantified by the same set of interfacial parameters.
\end{abstract}

A thorough understanding of charge and spin transport is crucial for the development of devices based on the electric control of spin currents.~\cite{zutic2004spintronics, Vignale2009}. In this respect, the charge-spin interconversion via spin-orbit coupling (SOC) plays a key role. SOC leads to the widely studied spin Hall (SHE)~\cite{sinova2015spin,valenzuela2006direct,kimura2007room} and Edelstein (EE) effects~\cite{aronov1989nuclear,edelstein1990spin,ando2017spin,soumyanarayanan2016emergent}, which are at the basis of spin-orbit torque memories~\cite{miron2011perpendicular,liu2012spin,safeer2016spin} and spin-based logic devices~\cite{pham2020spin,manipatruni2019scalable}.

Of particular interest are systems with sizable spin-charge interconversion at the interface between an insulator (I) that contains a heavy element, and a normal metal (N) with negligible SOC and long spin relaxation length, as for example 
BiO$_x$/Cu bilayers. In these systems, the spin-charge interconversion occurs at the hybrid interface via an interfacial spin-orbit coupling (ISOC)~\cite{karube2016experimental,kim2017evaluation}. Whereas the electronic transport in N is well described by customary drift-diffusion equations, the interfacial effects occur at atomic scales near the interface and, hence, their inclusion into the drift-diffusion model is kind of subtle. Some works use an intuitive picture based on the idealized two-dimensional Rashba model and assume the existence of a two-dimensional electron gas with Rashba SOC at the interface~\cite{kim2017evaluation,nakayama2016rashba,tsai2018clear}, in which the conversion takes place via the EE and its inverse (IEE). Such description is clearly valid for conductive surface states in (e.g.~topological) insulators~\cite{rojas2016spin,kondou2016fermi} or two-dimensional electron gases~\cite{lesne2016highly,vaz2019mapping}. However, in metallic systems, it requires the introduction of additional microscopic parameters to model the coupling between interface states and the diffusive motion of electrons in the metal. Moreover, the very existence of a well-defined two-dimensional interface band, and its relevance for the electronic transport, in systems like BiO$_x$/Cu is not obvious as realistic device structures are frequently polycristalline and quite disordered. Apparently, the physical picture involving the interface band with Rashba SOC is merely one of the possible interpretations for the experimental data. One can contemplate other microscopic scenarios to describe the charge-spin coupling in I/N systems. For example, in the BiO$_x$/Cu interface, one could assume that Bi atoms diffuse into Cu and induce an effective ``extrinsic'' SOC in a thin layer near the interface. Alternatively, a spin-charge interconversion can be generated via an interfacial spin-dependent scattering of the bulk Bloch states (see, for example, Refs.~\cite{LinYok2011,TokKraVig2015PRB,BorTok2017PRB}). Each of these models will provide a different set of microscopic parameters, which usually have to be inferred from the measurements of macroscopic transport properties.

In this Letter, we approach the problem from a different angle and propose a universal theoretical framework which is independent of microscopic mechanisms and details. Specifically, we provide the basic drift-diffusion theory describing the charge and spin transport in I/N structures. To account for the ISOC, we use the effective boundary conditions (BCs) derived in Ref.~\citen{borge2019boundary}. 
In this work, such BCs basically describe two types of interfacial processes mediated by ISOC: spin-charge interconversion and spin losses, quantified respectively by the interfacial spin-to-charge/charge-to-spin conductivities, $\sigma_{\rm{sc/cs}}$, and the spin loss conductances $G_{\rm{\parallel/\perp}}$ for spins polarized parallel/perpendicular to the interface. The efficiency of spin-charge interconversion, which is central to the development of spintronic devices, is determined by the ratio between the strengths of these two processes. 
This ratio coincides with the widely used conversion efficiency, the inverse Edelstein length $\lambda_{\rm{IEE}}$, such that $\lambda_{\rm{IEE}} = \sigma_{\rm{sc}}/G_{\parallel}$. We apply our theory to describe two typical experimental setups: a lateral spin valve (LSV) made of Cu wires with a middle wire partially covered by a BiO$_x$ layer (see Fig.~\ref{fig_LSV}a), where non-local resistances are measured, and a BiO$_x$/Cu/YIG trilayer Hall bar (see Fig.~\ref{fig_SMR}a), where we measure the spin Hall magnetoresistance (SMR). From contrasting the experimental results with our theory, we demonstrate that $\sigma_{\rm{sc}}=\sigma_{\rm{cs}}$, which confirms the Onsager reciprocity. In addition, both experiments can be described by similar values of the ISOC parameters {when measured at the same temperature}. This confirms that the spin-charge interconversion in those systems only depends on the intrinsic properties of the BiO$_x$/Cu interface. 

\section{Model and method}
\begin{figure*}[h!]
\includegraphics[scale=0.5]{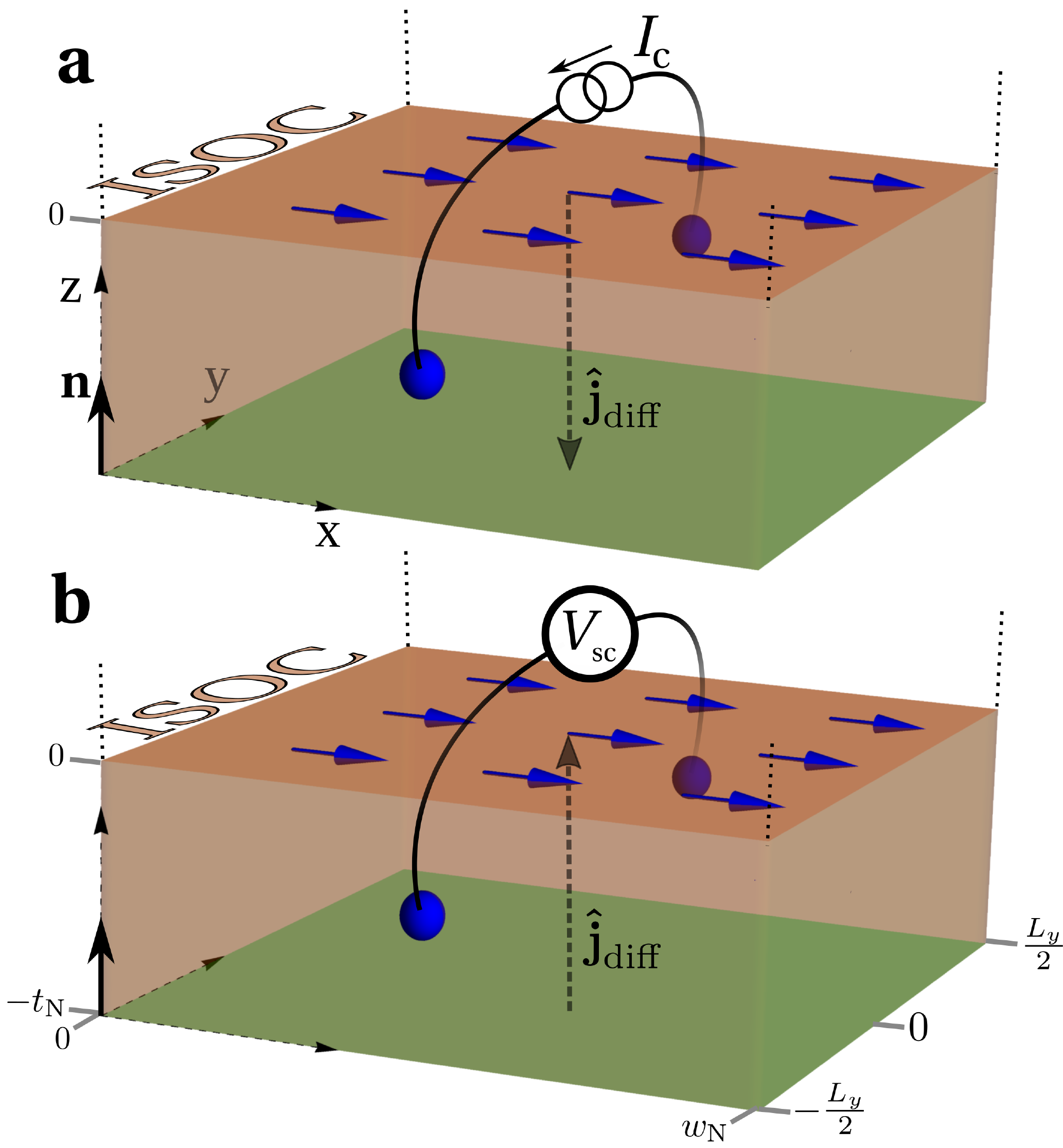}
\caption{\textbf{Sketch of the system under study}. A normal metal, at $z<0$, is adjacent to a non-magnetic insulator, at $z>0$. ISOC is finite at the interface with normal vector ${\bf{n}}$. (a) Charge-to-spin conversion: a charge current $I_{\rm{c}}$ induces a spin current density $\hat{\bf{j}}_{\rm{diff}}$. 
(b) Spin-to-charge conversion: a spin current density, $\hat{\bf{j}}_{\rm{diff}}$, induces at the interface a voltage drop perpendicular to the polarization of $\hat{\bf{j}}_{\rm{diff}}$.}
\label{fig_th_s}
\end{figure*}

We consider a customary setup consisting of a normal metal (N) with negligible spin-orbit coupling adjacent to a non-magnetic insulator (I), see Fig.~\ref{fig_th_s}. In the N layer, spin and charge transport is described by the diffusion equations: 
\begin{equation}\label{eq_deqs}
  \nabla^2 \hat{\mu} = \frac{ \hat{\mu}}{\lambda_{\rm{N}}^2}\; ,
\end{equation} 
\begin{equation}\label{eq_deqc}
  \nabla^2 \mu = 0\; .
\end{equation}
Here, $\hat{\mu}=(\mu^x,\mu^y,\mu^z)$ and $\mu$ are the spin and charge electrochemical potentials (ECP), where hat symbol indicates spin pseudovector. It is assumed that N has inversion symmetry and the spin relaxation is isotropic and described by the bulk spin diffusion length $\lambda_{\rm{N}}$~\cite{sanz2019nonlocal}. The diffusive charge and spin currents are defined as $e\hat{\bf{j}} = -\sigma_{\rm{N}} \nabla \hat{\mu}$ and $e{\bf{j}} = -\sigma_{\rm{N}} \nabla \mu$, respectively, with $e=-|e|$.

Equations~\eqref{eq_deqs} and \eqref{eq_deqc} are complemented by boundary conditions (BCs) at the interfaces. Specifically, at the interface with vacuum, one imposes a zero current condition. Whereas, at the I/N interface with ISOC, the BCs read~\cite{borge2019boundary}:
\begin{equation}\label{eq_BCBi_s}
  -\sigma_{\rm{N}} (\nabla \cdot {\bf{n}}) \hat{\mu} \big|_{0} = G_{\perp}\ \hat{\mu}_{\perp} \big|_{0} + G_{\parallel}\ \hat{\mu}_{\parallel} \big|_{0} + \sigma_{\rm{cs}} \left( {\bf{n}} \times \nabla \right) \mu \big|_{0} \; ,
\end{equation} 
\begin{equation}\label{eq_bc_cont}
  -\sigma_{\rm{N}}(\nabla \cdot {\bf{n}}) \mu \big|_{0} = \sigma_{\rm{sc}} \left( {\bf{n}} \times \nabla \right) \hat{\mu} \big|_{0} \; .
\end{equation} 
We denote by ${\bf{n}}$ the vector normal to the interface, pointing from the metal towards the interface, and $\sigma_{\rm{N}}$ is the conductivity of N. Equation \eqref{eq_BCBi_s} is the BC for the spin ECP. The last term in the r.h.s.~describes the charge-to-spin conversion, whose efficiency is described by the conductivity $\sigma_{\rm{cs}}$. As schematically shown in Fig.~\ref{fig_th_s}a, this term couples an effective electric field and the spin current density at the interface~\cite{amin2018interface,amin2016phenomenology,amin2016formalism,borge2019boundary}, and can be interpreted as an interfacial spin Hall conductivity. Alternatively, it can be interpreted as if the charge current induces an homogeneous spin ECP via an interfacial Edelstein effect (EE), which in turn diffuses into the bulk conductor. These two interpretations are fully compatible within the present formalism. The second type of processes which take place at the interface are spin losses (first two terms in the r.h.s.~of Eq.~\eqref{eq_BCBi_s}), which are quantified by the corresponding spin loss conductance density per area $G_{\rm{\perp/\parallel}}$ for spins perpendicular/parallel ($\hat{\mu}_\perp$/$\hat{\mu}_\parallel$) to the interface.

Equation~\eqref{eq_bc_cont} is the BC for the charge ECP. The charge is obviously conserved and, therefore, the r.h.s.~only contains the spin-to-charge conversion between a diffusive bulk spin current and an interfacial charge current, which induces a voltage drop. The last term in Eqs.~\eqref{eq_BCBi_s} and \eqref{eq_bc_cont} describes reciprocal effects.~\footnote{Symmetry arguments alone cannot fix the relation between $\sigma_{\rm{sc}}$ and $\sigma_{\rm{cs}}$~\cite{borge2019boundary}. However, we will see by contrasting theory with experiment, that reciprocity requires $\sigma_{\rm{sc}} = \sigma_{\rm{cs}}$.} Equation~\eqref{eq_bc_cont} can be interpreted as an interfacial inverse spin Hall effect (ISHE), which couples the charge and the spin currents. But again, an alternative  interpretation is possible: from the conservation of the charge current at the interface, we can relate the bulk charge current to the divergence of an interfacial current as $\sigma_{\rm{N}}(\nabla \cdot {\bf{n}}) \mu \big|_{0} = -e \nabla\cdot {\bf{j}}_{\rm{I}}$. Comparing this last equation with Eq.~\eqref{eq_bc_cont}, we may define the interfacial charge current density as:~\footnote{Notice that, in principle, an additional divergenceless term may appear in the r.h.s.~of Eq.~\eqref{eq_jI}. Indeed, as demonstrated in Ref.~\citen{borge2019boundary}, symmetry allows for a term proportional to the out-of-plane component of the spin ECP. In the present work, we only consider spin polarization parallel to the I/N interface and, hence, we neglect that term.}
\begin{equation}\label{eq_jI}
  e {\bf{j}}_{\rm{I}} = -\sigma_{\rm{sc}} \left( {\bf{n}} \times \hat{\mu} \right) \big|_{0} \; .
\end{equation} 
Written this way, BC~\eqref{eq_bc_cont} describes the conversion of a non-equilibrium spin into an interfacial charge current, which corresponds to what is usually called the interfacial inverse Edelstein effect (IEE), as illustrated in Fig.~\ref{fig_th_s}b.

Within this last interpretation, we can introduce the commonly used  conversion parameter $\lambda_{\rm{IEE}}$. It has dimensions of a length and is defined as the ratio between the amplitude of the induced interfacial charge current density, ${\bf{j}}_{\rm{I}}$, and the amplitude of the spin current injected to the interface from the bulk, $\sigma_{\rm{N}} (\nabla \cdot {\bf n}) \hat{\mu} \big|_{0}$. From Eq.~\eqref{eq_jI}, one can see that the effect is finite only if the spin current is polarized in a direction parallel to the interface. By using Eqs.~\eqref{eq_BCBi_s} and \eqref{eq_jI}, we then obtain:
\begin{equation}\label{eq_l_IEE}
\lambda_{\rm{IEE}} = \frac{\sigma_{\rm{sc}}}{G_{\parallel}} \; .
\end{equation}
This is a remarkable result that follows straightforwardly from our description. $\lambda_{\rm{IEE}}$ is purely determined by interfacial parameters and it is indeed a quantification of the conversion efficiency since it is the ratio between the spin-to-charge conversion and the spin loss at the interface. Note that both parameters, $\sigma_{\rm{sc}}$ and $G_\parallel$, are finite only in the presence of a finite ISOC. Their specific values depend on the microscopic properties of the interface, which are intrinsic for each material combination. Both $\sigma_{\rm{sc}}$ and $G_{\parallel}$ may depend on temperature and, thus, $\lambda_{\rm{IEE}}$ is also temperature dependent. 

From an experimental perspective, the spin-to-charge conversion is usually detected electrically, by measuring a voltage drop. Typical devices are shown in Figs.~\ref{fig_LSV}a and \ref{fig_SMR}a, where transverse voltages are measured. In theory, one needs to determine and integrate the charge ECP to determine the averaged voltage drop between two points. For concreteness, we consider the generic setup shown in Fig.~\ref{fig_th_s}b, in which a spin current polarized in the $x$ direction flows towards the interface, such that a voltage difference is generated in the transverse, $y$, direction according to Eq.~\eqref{eq_jI}. The averaged voltage drop is measured between the points $y = \pm L_y/2$, with $L_y$ the total length in the $y$ direction, and is given by (see Note~S1 for details):
\begin{equation}\label{eq_Vsc_gen}
  V_{\rm{sc}} = \frac{\sigma_{\rm{sc}}}{e\sigma_{\rm{N}}A_{\rm{N}}} \iint_{-\frac{L_y}{2}}^{\frac{L_y}{2}} \left( {\bf{n}} \times \hat{\mu} |_0 \right)\cdot{\bf{e}}_y\ dxdy \; ,
\end{equation}
where $A_{\rm{N}} = t_{\rm{N}} w_{\rm{N}}$ is the cross-sectional area of the wire, with $t_{\rm{N}}$ and $w_{\rm{N}}$ being its thickness and width, respectively, over which the voltage drop is averaged. Equation~\eqref{eq_Vsc_gen} shows that the voltage drop between two points is proportional to the total spin accumulation created via the ISOC between them.
 
In the following, we apply the above formalism to derive the voltage drop associated to spin-charge interconversions in two different devices.

\section*{Results and discussion}
\begin{figure*}[h!]
\includegraphics[scale=0.5]{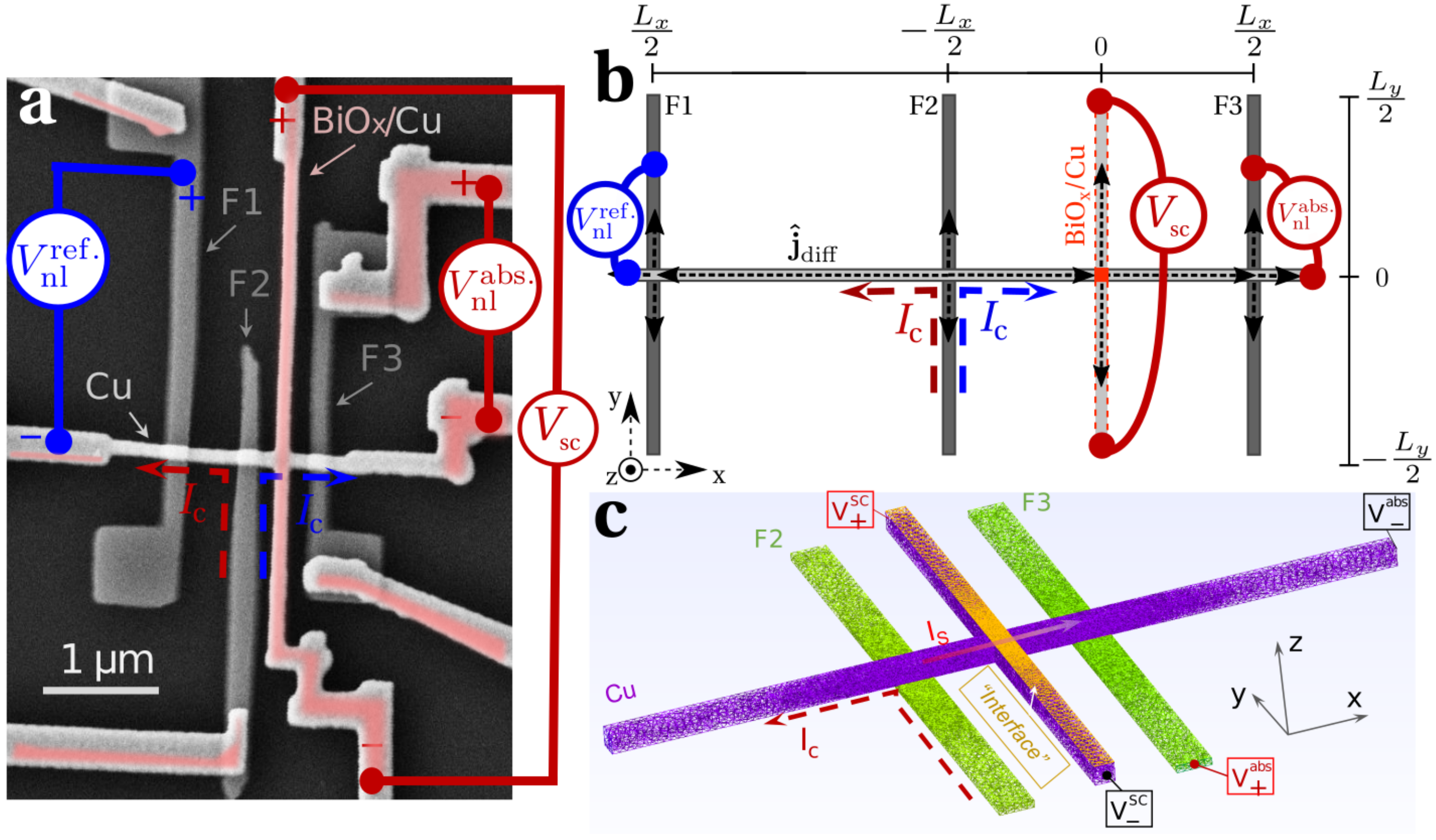}
\caption{\textbf{Lateral spin valve}. (a) SEM image of the two Py/Cu lateral spin valves. The reference LSV uses ferromagnets F1 and F2.
Non-local voltage $V_{\rm{nl}}^{\rm{ref}}$ is measured (blue circuit). 
The spin absorption experiment is performed in the LSV between F2 and F3, with a middle Cu wire covered with BiO$_x$ (light red covering). The non-local voltage 
$V_{\rm{nl}}^{\rm{abs.}}$ is measured (red circuit). 
In both cases the external magnetic field is applied along the $y$ axis.
The spin to charge conversion is detected by measurement of the transverse voltage $V_{\rm{sc}}$ after injection of a current from F2 (red circuit). In this case the external magnetic field is applied along the $x$ axis. 
(b) Effective one-dimensional model of the device. (c) Geometry and mesh of the 3D finite element method model used for simulating the spin absorption and spin-to-charge conversion.  The BiO$_x$/Cu interface is simulated as a thin layer (yellow) on top of the transverse Cu wire (purple). }
\label{fig_LSV}
\end{figure*}

We start analyzing the double LSV shown in Fig.~\ref{fig_LSV}a (see Note~S2 for fabrication and measurement details). A charge current $I_{\rm{c}}$ is injected from the Permalloy (Py) injector F2 into a Cu wire. F2 forms a LSV either with the detector F1 or F3. We use the LSV between F2 and F1 as a reference setup.  In the LSV between F2 and F3 there is an additional transverse Cu wire covered by a BiO$_x$ layer (red in Fig.~\ref{fig_LSV}a) and, hence, ISOC at the BiO$_x$/Cu interface. In this case, part of the spin current in the main Cu wire is absorbed and converted to a transverse charge current (see Fig.~\ref{fig_th_s}b).

Quantitative description of the spin injection, diffusion, and detection in LSVs has been widely studied in the literature~\cite{takahashi2003spin, niimi2014extrinsic}. In our setup, the thickness $t_{\rm{N}}$ and width $w_{\rm{N}}$ of the Cu wires are much smaller than the spin diffusion length $\lambda_{\rm{N}}$, and one can integrate the spin diffusion equation~\eqref{eq_deqs} over the wire cross-section and simplify it to a one-dimensional problem~\cite{niimi2014extrinsic,isshiki2019experimentally,takahashi2003spin}, as sketched in Fig.~\ref{fig_LSV}b. At the BiO$_x$/Cu middle wire, the $z$-integration using the BC at $z=0$ of Eq.~\eqref{eq_BCBi_s}, leads to a renormalization of $\lambda_{\rm{N}}$ (see Note~S3):
\begin{equation}\label{eq_lN||}
  \lambda_{\rm{N}\parallel} = \frac{\lambda_{\rm{N}}}{\sqrt{1 + \frac{G_{\parallel}\lambda_{\rm{N}}^2}{\sigma_{\rm{N}} t_{\rm{N}}}}} \; ,
\end{equation}
where we neglect corrections of order $\sigma_{\rm{sc}}^2$.

At the crossing point, $x=0$ in Fig.~\ref{fig_LSV}b, we use Kirchhoff's law for the spin currents (see Note~S3):
\begin{equation}\label{eq_BCnode}
-A_{\rm{N}}\sigma_{\rm{N}} \partial_x \hat{\mu}_{\parallel} \Big|_{0^-}^{0^+} = - G_{\rm{N}\parallel} \hat{\mu}_{\parallel} \Big|_{x=0} - A_{\rm{n}}^{\rm{eff}} \sigma_{\rm{cs}} \frac{ e j_{\rm{c}}}{\sigma_{\rm{N}}}\ \hat{\bf{e}}_x \; .
\end{equation}
Here, $G_{\rm{N}\parallel} = \frac{t_{\rm{N}} \sigma_{\rm{N}} A_{\rm{n}}^{\rm{eff}}}{\lambda_{\rm{N}\parallel}^2}$ is the effective spin (bulk) conductance of the BiO$_x$/Cu wire, with $A_{\rm{n}}^{\rm{eff}} = w_{\rm{N}} (w_{\rm{N}} + 2\lambda_{\rm{N}\parallel})$. The latter is the effective area of the BiO$_x$/Cu interface that absorbs (injects) spin current. Indeed, the r.h.s.~of this equation corresponds to Eq.~\eqref{eq_BCBi_s} with an effective spin loss conductance counting for both the interfacial and bulk spin losses at the middle wire. The last term in Eq.~\eqref{eq_BCnode} corresponds to the last term in Eq. \eqref{eq_BCBi_s} and it is proportional to the total injected charge current $I_{\rm{c}}$ along the middle wire oriented in the $y$ direction. If we assume an homogeneously distribution of the current, then ${\bf{j}}_{\rm{c}} = \frac{I_{\rm{c}}}{A_{\rm{N}}} \hat{\bf{e}}_y$. 

The Cu/Py interfaces are described by the following BC~\cite{pham2016ferro,takahashi2003spin,kimura2005estimation}:
\begin{equation}\label{eq_BCFe_s}
  \begin{array}{c}
     - A_{\rm{N}} \sigma_{\rm{N}} \partial_x \hat{\mu}_{\parallel} \Big|^{x = - \frac{L_x^-}{2}}_{x = - \frac{L_x^+}{2}} = - A_{\rm{F}} \left( \sigma^*_{\rm{F}} \partial_y \hat{\mu}_{\parallel {\rm{F}}2} \big|_{y = 0} + p_{\rm{F}} e j_{\rm{c}} \right) \;, \\
     - A_{\rm{N}} \sigma_{\rm{N}} \partial_x \hat{\mu}_{\parallel} \Big|^{x = \frac{L_x^-}{2}}_{x = \frac{L_x^+}{2}} = - A_{\rm{F}} \sigma^*_{\rm{F}} \partial_y \hat{\mu}_{\parallel {\rm{F}}3} \big|_{y = 0} \;, 
  \end{array}
\end{equation} 
where $\hat{\mu}_{\parallel {\rm{F2}}}$ and $\hat{\mu}_{\parallel {\rm{F3}}}$ are the spin ECPs at ferromagnets F2 and F3, respectively. The polarization and effective conductivity at the ferromagnets are $p_{\rm{F}}$ and $\sigma^*_{\rm{F}} = \sigma_{\rm{F}} (1 - p_{\rm{F}}^2)$, respectively, and $L_x$ is the distance between consecutive ferromagnetic wires. Since the spin current strongly decreases at the ferromagnets, $A_{\rm{F}}$ corresponds to the Py/Cu junction area~\cite{takahashi2003spin}. The last term of the first equation above is proportional to the charge current density, $j_{\rm{c}} = \frac{I_{\rm{c}}}{A_{\rm{F}}}$, (homogeneously) injected at the ferromagnetic wire. For the description of the reference LSV, we substitute F3 by F1 in the second BC. Because the Py/Cu interfaces are good metallic contacts we assume the continuity of $\hat{\mu}_{\parallel}$. This condition, together with the one-dimensional version of Eq.~\eqref{eq_deqs} and the BCs~\eqref{eq_BCnode} and \eqref{eq_BCFe_s}, determine the full spatial dependence of $\hat{\mu}_{\parallel}$.

\begin{figure*}[h!]
\includegraphics[scale=0.5]{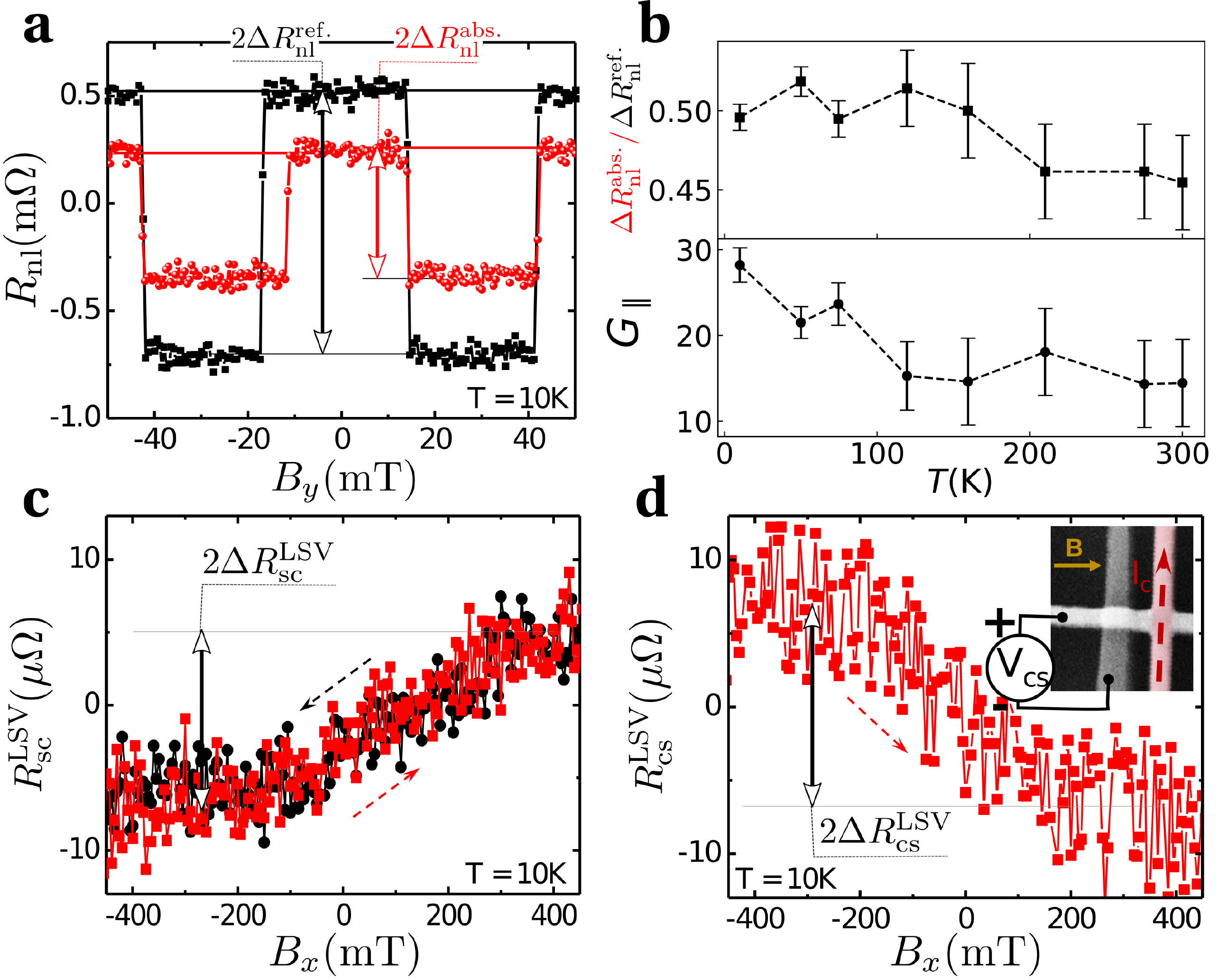}
\caption{{\bf{LSV experimental results.}} (a) Non-local resistance as a function of $B_y$ (trace and retrace) measured at $I_{\rm{c}} = 70 \mu{\rm{A}}$ and $10$ K for the reference LSV (black squares) and the LSV with the middle BiO$_x$/Cu wire (red circles). From this measurement, we extract $\Delta R_{\rm{nl}}^{\rm{ref.}}$ and $\Delta R_{\rm{nl}}^{\rm{abs.}}$. (b) Upper panel: the spin absorption ratio ($\Delta R_{\rm{nl}}^{\rm{abs.}}/\Delta R_{\rm{nl}}^{\rm{ref.}}$) as a function of temperature; lower panel, the corresponding interfacial spin-loss conductance in units of $(\Omega^{-1}\mu {\rm{m}}^{-2})$ as a function of the temperature, calculated from Eq.~\eqref{eq_Rnl}. (c) Non-local resistance as a function of $B_x$ (trace and retrace) measured at $I_{\rm{c}} = 70 \mu{\rm{A}}$ and $10\ {\rm{K}}$. Each curve is an average of 7 sweeps. The spin-to-charge conversion signal ($2\Delta R_{\rm{sc}}^{\rm{LSV}}$) is extracted. (d) Non-local resistance as a function of $B_x$ (trace only) measured at $I_{\rm{c}} = 150 \mu{\rm{A}}$ and $10$ K using the configuration of the inset, which is reciprocal to the one of panel (c). From this measurement we extract the charge-to-spin conversion signal ($2\Delta R_{\rm{cs}}^{\rm{LSV}}$). The curve is an average of 4 sweeps.}
\label{fig_LSVexp}
\end{figure*}

We are interested in the value of $\hat{\mu}_{\parallel}$ at the detectors F1/F3 
which is proportional to the non-local voltages $V_{\rm{nl}} = e^{-1}p_{\rm{F}} \hat{\mu}_{\parallel {\rm{F}1/\rm{F}3}}\big|_{y=0}$~\cite{takahashi2003spin,kimura2004spin} (see Fig.~\ref{fig_LSV}b). From such non-local measurement, we determine the non-local resistance, $R_{\rm{nl}} = V_{\rm{nl}} / I_{\rm{c}}$, where $I_{\rm{c}}$ is the current injected from F2. 
The value of $R_{\rm{nl}}$ changes sign when the magnetic configuration of injector and detector ferromagnets changes from parallel, $R^{\rm{P}}_{\rm{nl}}$, to antiparallel, $R^{\rm{AP}}_{\rm{nl}}$, which experimentally allows us to remove any baseline resistance coming from non-spin related effects by taking $\Delta R_{\rm{nl}}= R^{\rm{P}}_{\rm{nl}} -R^{\rm{AP}}_{\rm{nl}}$ (see Fig.~\ref{fig_LSVexp}a). 
Comparing the non-local resistance measured at F3, $\Delta R^{\rm{abs}}_{\rm{nl}}$, with the one measured at F1 at the reference LSV, $\Delta R^{\rm{ref}}_{\rm{nl}}$, the magnitude of the spin absorption~\cite{niimi2014extrinsic,isasa2016origin} and, therefore, the value of the spin loss conductance, $G_{\parallel}$, can be obtained from our model. For this, we compute the ratio $\Delta R^{\rm{abs.}}_{\rm{nl}}/\Delta R^{\rm{ref.}}_{\rm{nl}} = \hat{\mu}_{\parallel {\rm{F}3}}/\hat{\mu}_{\parallel {\rm{F}1}} |_{y=0}$ by solving the full boundary problem and obtain the following expression:
\begin{equation}\label{eq_Rnl}
\frac{\Delta R^{\rm{abs.}}_{\rm{nl}}}{\Delta R^{\rm{ref.}}_{\rm{nl}}} = \left[ 1 + \frac{G_{\rm{N}\parallel}}{2G_{\rm{N}}}\frac{(G_{\rm{F}} + 2G_{\rm{N}}) - G_{\rm{F}}\ e^{-\frac{L_x}{\lambda_{\rm{N}}}}}{(G_{\rm{F}} + 2G_{\rm{N}}) + G_{\rm{F}}\ e^{-\frac{L_x}{\lambda_{\rm{N}}}}} \right]^{-1} \; .
\end{equation}
Here, $G_{i} = \frac{\sigma_i A_i}{\lambda_i}$ are the spin conductances of the Cu and Py wires, with $\lambda_i$, $\sigma_i$, and $A_i$, the corresponding parameters for the bare Cu wire ($i = \rm{N}$) and the ferromagnet ($i$ = F). The form of Eq.~\eqref{eq_Rnl} agrees with the one obtained in previous works~\cite{isasa2016origin}. However, our formulation is more general since it distinguishes via $G_{\rm{N}\parallel}$ between interfacial and bulk losses at the BiO$_x$/Cu wire. Consequently, we can ensure that our calculation of the ISOC parameter $G_{\parallel}$ and, therefore, $\lambda_{\rm{IEE}}$, is only related to interfacial effects (see Eqs.~\eqref{eq_l_IEE}, \eqref{eq_lN||}, and \eqref{eq_BCnode}).

Interestingly, Fig.~\ref{fig_LSVexp}b shows weak temperature dependence of the absorption ratio, $\Delta R^{\rm{abs.}}_{\rm{nl}} / \Delta R^{\rm{ref.}}_{\rm{nl}}\approx 0.5$, revealing that about half of the spin current is absorbed at the BiO$_x$/Cu middle wire. For the calculation of $G_{\parallel}$ via Eq.~\eqref{eq_Rnl}, the resistivity of the Cu layer is carefully measured as a function of the temperature (see Note~S4), with $t_{\rm{N}} = w_{\rm{N}} = 80\ \operatorname{nm}$, and $L_x = 570\ {\rm{nm}}$. Assuming that the magnetic properties of Py and the specific spin resistivities of Py and Cu of our device are the same as in Ref.~\citen{sagasta2017spin}, we use the same temperature dependence of $\rho_{\rm{F}}$ and $p_{\rm{F}}$, and the constant spin resistivities $\lambda_{\rm{F}}/\sigma_{\rm{F}} = 0.91\ {\rm{f}}\Omega{\rm{m}}^2$ and $\lambda_{\rm{N}}/\sigma_{\rm{N}} = 18.3\ {\rm{f}}\Omega{\rm{m}}^2$. By inserting these experimental values in Eq.~\eqref{eq_Rnl} for different temperatures, we obtain the temperature dependence of the spin loss conductance $G_{\parallel}$ for the BiO$_x$/Cu interface shown in Fig.~\ref{fig_LSVexp}b. A slight decrease of $G_{\parallel}$ can be observed with increasing temperature, which seems to arise from the Cu conductivity. A linear relation between $G_{\parallel}$ and $\sigma_{\rm{N}}$ (see Note~S5A) suggests a Dyakonov-Perel mechanism of the spin loss, expected for a Rashba interface, which also agrees with the observations of Ref.~\citen{tsai2019enhanced}.

In addition, we can determine $\sigma_{\rm{sc/cs}}$ in the same device. By injecting a charge current $I_{\rm{c}}$ from F2, a $x$-polarized spin current is created and reaches the BiO$_x$/Cu wire, where a conversion to a transverse charge current occurs via Eq.~\eqref{eq_jI}. This is detected as a non-local voltage $V_{\rm{sc}}$ along the BiO$_x$/Cu wire and the non-local resistance $R_{\rm{sc}}^{\rm{LSV}} = V_{\rm{sc}}/I_{\rm{c}}$ is determined as a function of an in-plane magnetic field along the hard axis of F2, $B_x$. By reversing the orientation of the magnetic field, the opposite $R_{\rm{sc}}^{\rm{LSV}}$ is obtained, since the Py magnetization is reversed as well as the orientation of the spin polarization (see Fig. \ref{fig_LSVexp}c). The difference of the two values for $R_{\rm{sc}}^{\rm{LSV}}$, denoted as $2\Delta R_{\rm{sc}}^{\rm{LSV}}$ in Fig. \ref{fig_LSVexp}c, allows to remove any baseline resistance. By swapping the voltage and current probes, the reciprocal charge-to-spin conversion signal, $R_{\rm{cs}}^{\rm{LSV}} = V_{\rm{cs}}/I_{\rm{c}}$, can also be measured. 

Theoretically, from the calculation of the spatial dependence of the spin ECP, we compute both $V_{\rm{sc}}$, from Eq.~\eqref{eq_Vsc_gen}, and $V_{\rm{cs}} = e^{-1}p_{\rm{F}} \mu^x_{\parallel {\rm{F}2}}\big|_{y=0}$, by assuming a homogeneous spin absorption/injection at $A_{\rm{n}}^{\rm{eff}}$.
We obtain the same expressions, with opposite sign:
\begin{equation}\label{eq_Rsc_LSV}
\Delta R_{\rm{sc/cs}}^{\rm{LSV}} = \pm \frac{\sigma_{\rm{sc/cs}}}{\sigma_{\rm{N}}} \frac{A_{\rm{n}}^{\rm{eff}}}{A_{\rm{N}}} \frac{p_{\rm{F}}\ e^{\frac{L_x}{2\lambda_{\rm{N}}}} }{ G_{\rm{F}} \left( 1 - \frac{G_{\rm{N}\parallel}}{2G_{\rm{N}}} \right) + e^{\frac{L_x}{\lambda_{\rm{N}}}} (G_{\rm{F}} + 2G_{\rm{N}}) \left( 1 + \frac{G_{\rm{N}\parallel}}{2G_{\rm{N}}} \right) } \; .
\end{equation}
Experimentally, Figs.~\ref{fig_LSVexp}c and \ref{fig_LSVexp}d confirm the reciprocity between both measurements, $\Delta R_{\rm{cs}}^{\rm{LSV}} = \Delta R_{\rm{sc}}^{\rm{LSV}}$. The broken time reversal symmetry, due to the magnetic contacts, leads to the opposite sign for reciprocal measurements. By contrasting this with the result of Eq.~\eqref{eq_Rsc_LSV}, one confirms that $\sigma_{\rm{sc}} = \sigma_{\rm{cs}}$. 
The experimental value for $2\Delta R_{\rm{sc}}^{\rm{LSV}}\approx 15 \pm 3 \mu\Omega$ at 10 K yields $\sigma_{\rm{sc/cs}} \approx 44 \pm 9\ \Omega^{-1}\operatorname{cm}^{-1}$ for the spin-charge conductivity and $\lambda_{\rm{IEE}}\approx 0.16 \pm 0.03\operatorname{nm}$. 
This value is of same order of magnitude, but somewhat smaller, than previous reported results obtained by spin pumping experiments, $\lambda_{\rm{IEE}} \approx 0.2-0.7\operatorname{nm}$~\cite{karube2016experimental,tsai2018clear,tsai2019enhanced}, and LSV experiments, $\lambda_{\rm{IEE}} \approx 0.5-1\operatorname{nm}$~\cite{isshiki2019experimentally}. This discrepancy might be due to a different quality of the BiO$_x$/Cu interface: \textit{ex-situ} deposition in this experiment (see Note~S2) and \textit{in-situ} deposition in the other works.
The temperature dependence of the different parameters is presented in Note~S5B. One observes a decreasing trend of $\sigma_{\rm{sc}}$ by increasing the temperature, which translates in a decreasing of the Edelstein length, in agreement with previous literature~\cite{tsai2019enhanced}.

In order to check the accuracy of our 1D model, we have performed a 3D finite element method simulation detailed in Note~S6. Figure~\ref{fig_LSV}c shows the geometry of the simulated device and the mesh of the finite elements. The interface with ISOC is simulated as a thin layer with finite thickness $t_{\rm{int}}$ and characterized by a spin diffusion length $\lambda_{\rm{int}}$ and an effective spin Hall angle $\theta_{\rm{int}}^{\rm{eff}}$. Using the definition of the Edelstein length as $\frac{1}{2}\theta_{\rm{int}}^{\rm{eff}} t_{\rm{int}}$~\cite{sanchez2013spin}, we obtain $\lambda_{\rm{IEE}} = 0.10 \pm 0.02$ nm, in good agreement with the $\lambda_{\rm{IEE}}$ estimated from our 1D model.

\begin{figure*}[h!]
\includegraphics[scale=0.5]{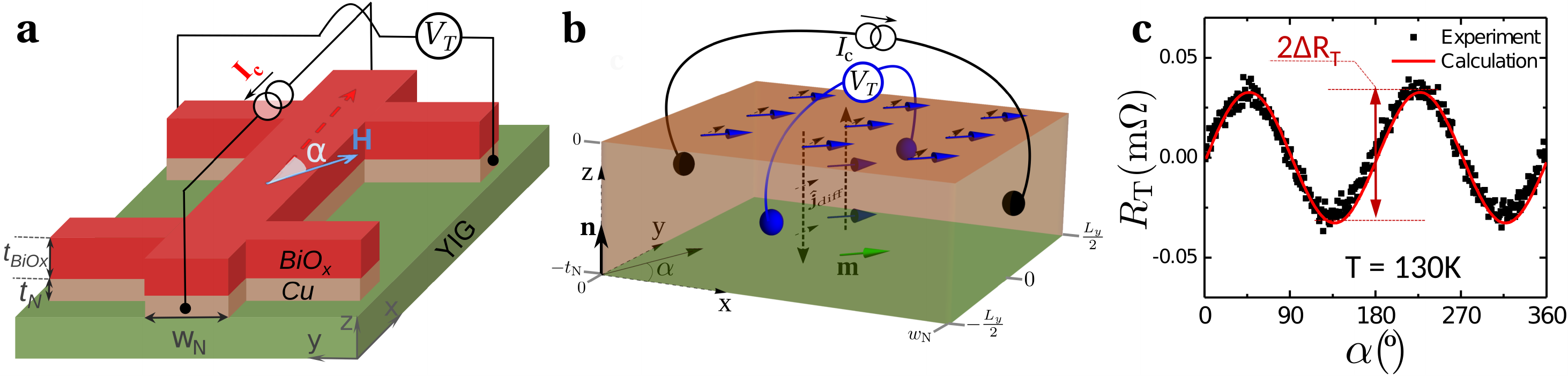}
\caption{\textbf{Spin magnetoresistance in multilayer device.} (a) Measurement configuration of the TADMR in BiO$_x$/Cu Hall-cross device on YIG. A charge current $I_{\rm{c}}$ is applied along the $x$ direction. The external H-field (100 mT) is applied in-plane the YIG substrate, $x$--$y$ plane, to drive the magnetisation of YIG. A voltmeter is applied to probe the transverse potential change $V_{\rm{T}}$ under open circuit conditions. $\alpha$ is the angle between the applied current vector (red dashed arrow) and the applied field (blue solid arrow) in the $x$--$y$ plane. (b) Sketch of the double spin-charge interconversion at the BiO$_x$/Cu interface with ISOC. First, charge current $I_{\rm{c}}$ injection induces a bulk $y$-polarized spin current density flowing towards the Cu/YIG interface, where it is back-reflected with mixed $x$ and $y$ polarizations. Secondly, the $x$-polarized contribution to the spin ECP at the ISOC interface, generates a voltage drop along the $y$ direction. (c) TADMR ($R_{\rm{T}}$) as a function of $\alpha$. $R_{\rm{T}}$ is the value of the measured transverse voltage divided by the applied current $I_{\rm{c}}$ (black squares). Solid curve corresponds to $R_{\rm{T}} = 0.03\sin(2\alpha)$.}
\label{fig_SMR}
\end{figure*}

In order to verify that both ISOC parameters, $G_{\parallel}$ and $\sigma_{\rm{sc}}$, are interface specific, we carry out an additional experiment involving a BiO$_x$/Cu interface. 
Namely, we measure the SMR in a Cu layer sandwiched between a BiO$_x$ (at $z=0$) and a YIG layers (at $z=-t_{\rm{N}}$), shaped as a Hall bar, as shown in Fig.~\ref{fig_SMR}a (see Note~S2 for fabrication and measurement details). In this setup, see Fig.~\ref{fig_SMR}b, a charge current $I_{\rm{c}}$ in the longitudinal direction ($x$ direction), 
%with a charge current density $j_c \approx I_{\rm{c}}/ (L_y t_{\rm{N}})$
induces an out-of-plane spin current density via the ISOC, described by the last term of Eq.~\eqref{eq_BCBi_s}. This spin current, polarized in the $y$ direction, propagates towards the Cu/YIG interface where is partly reflected and modified~\cite{chen2013theory,nakayama2013spin,zhang2019theory}. The reflected spin current diffuses back to the BiO$_x$/Cu interface, where it is converted back to an interfacial charge current via the reciprocal effect. Therefore, the overall effect is of second order in ISOC and proportional to $\sigma_{\rm{cs}}\sigma_{\rm{sc}} = \sigma^2_{\rm{sc}}$.

YIG is an insulating ferrimagnetic material and the electron spin reflection at the Cu/YIG interface depends on the direction of magnetization of the YIG, denoted as ${\bf m}$. The effective BC describing this interface is well-known and reads as follows:~\cite{brataas2000finite}~\footnote{Following our convention the vector ${\bf{n}}$ normal to the interface, points from the Cu towards the insulating layer.} 
\begin{equation}\label{eq_BCYIG_s}
  -\sigma_{\rm{N}} (\nabla \cdot {\bf{n}}) \hat{\mu} \big|_{-t_{\rm{N}}} = G_{\rm{s}}\ \hat{\mu} \big|_{-t_{\rm{N}}} + G_{\rm{r}}\ {\bf{m}} \times \left( \hat{\mu} \times {\bf{m}} \right) \big|_{-t_{\rm{N}}} + G_{\rm{i}}\ \left( {\bf{m}} \times \hat{\mu} \right) \big|_{-t_{\rm{N}}} \; .
\end{equation} 
Here, $G_{\rm{r,i}}$ are the real and imaginary parts of the spin-mixing conductance (per area), $G_{\uparrow\downarrow} = G_{\rm{r}} + i G_{\rm{i}}$, and $G_{\rm{s}}$ is the so-called spin-sink conductance. The values of these parameters are known for YIG, where $G_{\rm{i}} \ll G_{\rm{s}} < G_{\rm{r}}$ and, hence, $G_{\rm{i}}$ can be neglected~\cite{vlietstra2013exchange,althammer2013quantitative,kosub2018anomalous,zhang2019theory}. 

In the experiment, the transverse angular dependent magnetoresistance (TADMR) measurement is performed in the Hall bar of BiO$_x$/Cu grown on a YIG substrate as shown in Fig.~\ref{fig_SMR}a. The transverse voltage depends on the direction of the in-plane applied magnetic field, parameterized by the angle $\alpha$. The experimental results for the TADMR, $R_{\rm{T}} = V_{\rm{T}}/I_{\rm{c}}$,  are shown in Fig.~\ref{fig_SMR}c.
%with $I_{\rm{c}}$ being the applied current. 

To calculate $R_{\rm{T}}$, we first determine the spatial dependence of the spin ECP from Eq.~\eqref{eq_deqs} and BCs Eq.~\eqref{eq_BCBi_s} and Eq.~\eqref{eq_BCYIG_s} at the BiO$_x$/Cu and Cu/YIG interfaces, respectively. We assume that the system is translational invariant in the $x$--$y$ plane and reduce the diffusion problem to a 1D problem in the $z$ direction. We then use Eq.~\eqref{eq_Vsc_gen} to obtain $V_T$, averaged in the cross-sectional area $A_{\rm{N}}$. This results in: 
\begin{equation}\label{eq_Rsc_SMR}
R_{\rm{T}} \approx \frac{\sigma^2_{\rm{sc}}}{ 2\sigma^2_{\rm{N}} t^2_{\rm{N}}} \frac{ G_{\rm{r}} }{ ({ G_{\parallel} + G_{\rm{s}} })(G_{\parallel} + G_{\rm{s}} + G_{\rm{r}}) } \sin (2\alpha) = \Delta R_{\rm{T}} \sin (2\alpha)\; .
\end{equation}
We denote by $\Delta R_{\rm{T}}$ the amplitude of the modulation and assume that $\lambda_{\rm{N}}\gg t_{\rm{N}}$ (see Note~S7 for the exact expression). Note that the parameters of the Cu/YIG interface, $G_{\rm{r,s}}$, add to the spin loss at the BiO$_x$/Cu interface $G_{\parallel}$. We identify by comparison of Eqs.~\eqref{eq_BCBi_s} and \eqref{eq_BCYIG_s} two effective spin loss conductances, $G_x = (G_{\parallel} + G_{\rm{s}})$ and $G_y = (G_{\parallel} + G_{\rm{s}} + G_{\rm{r}})$, for spins polarized in the $x$ and $y$ directions, respectively. The amplitude of the SMR signal is then, according to Eq.~\eqref{eq_Rsc_SMR}, proportional to the difference $G_x-G_y$.

From Figure~\ref{fig_SMR}c, we estimate $\Delta R_{\rm{T}} \approx 0.03\ {\rm{m}}\Omega$ at $T = 130$ K. At this temperature, from the LSV measurements, we obtain for the BiO$_x$/Cu ISOC parameters $G_{\parallel} \approx 1.5 \times 10^{13} \Omega^{-1} \operatorname{m^{-2}}$ and $\sigma_{\rm{sc/cs}} \approx 11.3\ \Omega^{-1}\operatorname{cm}^{-1}$, as shown in Figs.~\ref{fig_LSVexp}b and S3b, respectively. 
The spin conductances $G_{\rm{s}}$ and $G_{\rm{r}}$ in light metal/YIG interfaces have been estimated in evaporated Cu and Al~\cite{das2019temperature,villamor2015modulation}. Whereas $G_{\rm{s}} = 3.6 \times 10^{12} \Omega^{-1} \operatorname{m^{-2}}$ for Cu/YIG~\cite{villamor2015modulation} is a consistent value in the literature~\cite{das2019temperature,cornelissen2016magnon}, the reported $G_{\rm{r}}$ is very low~\cite{villamor2015modulation}, as generally observed in evaporated metals on YIG~\cite{das2019temperature,vlietstra2013spin}. By substituting $G_{\parallel}$, $G_{\rm{s}}$, and $\sigma_{\rm{sc/cs}}$ values in Eq.~\eqref{eq_Rsc_SMR}, we obtain $G_{\rm{r}} \approx 6.1\times 10^{13} \Omega^{-1} \operatorname{m^{-2}}$. This value for sputtered Cu on YIG is much larger than that estimated in evaporated Cu on YIG, in agreement with the reported difference between sputtered and evaporated Pt~\cite{vlietstra2013spin}. Importantly, the obtained $G_{\rm{r}}$ satisfies the required condition $G_{\rm{s}} < G_{\rm{r}}$~\cite{zhang2019theory,cornelissen2016magnon} which confirms the validity of our estimation. 

\section*{Conclusions}
We present a complete and novel theoretical framework based on the drift-diffusion equations to accurately describe electronic transport in systems with ISOC at non-magnetic metal/insulator interfaces. Within our model, the interface is described by two type of processes: spin losses, parameterized by the interfacial conductances $G_{\parallel/\perp}$, and spin-charge interconversions, quantified by $\sigma_{\rm{sc}}$ and $\sigma_{\rm{cs}}$. These parameters are material-specific. The efficiency of the interconversion is quantified by the ratio $\sigma_{\rm{sc}}/G_{\parallel}$, which coincides with the commonly used Edelstein length $\lambda_{\rm{IEE}}$. The Onsager reciprocity is directly captured by $\sigma_{\rm{sc}}=\sigma_{\rm{cs}}$, as demonstrated by comparing our theoretical and experimental results. Our theory is an effective tool for an accurate quantification of spin-charge interconversion phenomena at interfaces, which is of paramount importance in many novel spintronic devices.

\begin{acknowledgement}
C.S-F, F.S.B., and I.V.T acknowledge funding by the Spanish Ministerio de Ciencia, Innovación y Universidades (MICINN) (Projects No. FIS2016-79464-P and No. FIS2017-82804-P), by Grupos Consolidados UPV/EHU del Gobierno Vasco (Grant No. IT1249- 19). The work of F.S.B. is partially funded by EU’s Horizon 2020 research and innovation program under Grant Agreement No. 800923 (SUPERTED). The work at nanoGUNE is supported by Intel Corporation through the Semiconductor Research Corporation under MSR-INTEL TASK 2017-IN-2744 and the ``FEINMAN'' Intel Science Technology Center, and by the Spanish MICINN under the Maria de Maeztu Units of Excellence Programme (MDM-2016-0618) and under project number MAT2015-65159-R and RTI2018-094861-B-100. V.T.P. acknowledges postdoctoral fellowship support from ``Juan de la Cierva--Formación'' program by the Spanish MICINN (grant numbers FJCI-2017-34494). E.S. thanks the Spanish MECD for a PhD fellowship (grant number FPU14/03102).
\end{acknowledgement}

\begin{suppinfo}
\begin{itemize}
  \item Additional details on the derivation of the spin-to-charge averaged voltage, Eq.~\eqref{eq_Vsc_gen}, and the renormalized spin diffusion length and node boundary condition for the LSV, Eqs.~\eqref{eq_lN||} and \eqref{eq_BCnode}, respectively; measured temperature dependence of the Cu resistivity and analysis on the temperature dependence of the ISOC parameters in the LSV; brief explanation of the 3D simulation and the relation between the simulation and ISOC parameters; theoretical result for the transverse resistance measured in the multilayer Hall bar, i.e., which leads to Eq.~\eqref{eq_Rsc_SMR}. The experimental details of the nanofabrication and measurements of the LSV and multilayer Hall bar devices are also included.
\end{itemize}
\end{suppinfo}

\bibliography{Bibl}

\providecommand{\latin}[1]{#1}
\makeatletter
\providecommand{\doi}
  {\begingroup\let\do\@makeother\dospecials
  \catcode`\{=1 \catcode`\}=2 \doi@aux}
\providecommand{\doi@aux}[1]{\endgroup\texttt{#1}}
\makeatother
\providecommand*\mcitethebibliography{\thebibliography}
\csname @ifundefined\endcsname{endmcitethebibliography}
  {\let\endmcitethebibliography\endthebibliography}{}
\begin{mcitethebibliography}{52}
\providecommand*\natexlab[1]{#1}
\providecommand*\mciteSetBstSublistMode[1]{}
\providecommand*\mciteSetBstMaxWidthForm[2]{}
\providecommand*\mciteBstWouldAddEndPuncttrue
  {\def\EndOfBibitem{\unskip.}}
\providecommand*\mciteBstWouldAddEndPunctfalse
  {\let\EndOfBibitem\relax}
\providecommand*\mciteSetBstMidEndSepPunct[3]{}
\providecommand*\mciteSetBstSublistLabelBeginEnd[3]{}
\providecommand*\EndOfBibitem{}
\mciteSetBstSublistMode{f}
\mciteSetBstMaxWidthForm{subitem}{(\alph{mcitesubitemcount})}
\mciteSetBstSublistLabelBeginEnd
  {\mcitemaxwidthsubitemform\space}
  {\relax}
  {\relax}

\bibitem[\ifmmode \check{Z}\else \v{Z}\fi{}uti\ifmmode~\acute{c}\else
  \'{c}\fi{} \latin{et~al.}(2004)\ifmmode \check{Z}\else
  \v{Z}\fi{}uti\ifmmode~\acute{c}\else \'{c}\fi{}, Fabian, and
  Das~Sarma]{zutic2004spintronics}
\ifmmode \check{Z}\else \v{Z}\fi{}uti\ifmmode~\acute{c}\else \'{c}\fi{},~I.;
  Fabian,~J.; Das~Sarma,~S. Spintronics: Fundamentals and applications.
  \emph{Rev. Mod. Phys.} \textbf{2004}, \emph{76}, 323--410\relax
\mciteBstWouldAddEndPuncttrue
\mciteSetBstMidEndSepPunct{\mcitedefaultmidpunct}
{\mcitedefaultendpunct}{\mcitedefaultseppunct}\relax
\EndOfBibitem
\bibitem[Vignale(2009)]{Vignale2009}
Vignale,~G. Ten Years of Spin Hall Effect. \emph{Journal of Superconductivity
  and Novel Magnetism} \textbf{2009}, \emph{23}, 3\relax
\mciteBstWouldAddEndPuncttrue
\mciteSetBstMidEndSepPunct{\mcitedefaultmidpunct}
{\mcitedefaultendpunct}{\mcitedefaultseppunct}\relax
\EndOfBibitem
\bibitem[Sinova \latin{et~al.}(2015)Sinova, Valenzuela, Wunderlich, Back, and
  Jungwirth]{sinova2015spin}
Sinova,~J.; Valenzuela,~S.~O.; Wunderlich,~J.; Back,~C.~H.; Jungwirth,~T. Spin
  Hall effects. \emph{Rev. Mod. Phys.} \textbf{2015}, \emph{87},
  1213--1260\relax
\mciteBstWouldAddEndPuncttrue
\mciteSetBstMidEndSepPunct{\mcitedefaultmidpunct}
{\mcitedefaultendpunct}{\mcitedefaultseppunct}\relax
\EndOfBibitem
\bibitem[Valenzuela and Tinkham(2006)Valenzuela, and
  Tinkham]{valenzuela2006direct}
Valenzuela,~S.~O.; Tinkham,~M. Direct electronic measurement of the spin Hall
  effect. \emph{Nature} \textbf{2006}, \emph{442}, 176--179\relax
\mciteBstWouldAddEndPuncttrue
\mciteSetBstMidEndSepPunct{\mcitedefaultmidpunct}
{\mcitedefaultendpunct}{\mcitedefaultseppunct}\relax
\EndOfBibitem
\bibitem[Kimura \latin{et~al.}(2007)Kimura, Otani, Sato, Takahashi, and
  Maekawa]{kimura2007room}
Kimura,~T.; Otani,~Y.; Sato,~T.; Takahashi,~S.; Maekawa,~S. Room-Temperature
  Reversible Spin Hall Effect. \emph{Phys. Rev. Lett.} \textbf{2007},
  \emph{98}, 156601\relax
\mciteBstWouldAddEndPuncttrue
\mciteSetBstMidEndSepPunct{\mcitedefaultmidpunct}
{\mcitedefaultendpunct}{\mcitedefaultseppunct}\relax
\EndOfBibitem
\bibitem[Aronov and Lyanda-Geller(1989)Aronov, and
  Lyanda-Geller]{aronov1989nuclear}
Aronov,~A.; Lyanda-Geller,~Y.~B. Nuclear electric resonance and orientation of
  carrier spins by an electric field. \emph{Soviet Journal of Experimental and
  Theoretical Physics Letters} \textbf{1989}, \emph{50}, 431\relax
\mciteBstWouldAddEndPuncttrue
\mciteSetBstMidEndSepPunct{\mcitedefaultmidpunct}
{\mcitedefaultendpunct}{\mcitedefaultseppunct}\relax
\EndOfBibitem
\bibitem[Edelstein(1990)]{edelstein1990spin}
Edelstein,~V.~M. Spin polarization of conduction electrons induced by electric
  current in two-dimensional asymmetric electron systems. \emph{Solid State
  Communications} \textbf{1990}, \emph{73}, 233--235\relax
\mciteBstWouldAddEndPuncttrue
\mciteSetBstMidEndSepPunct{\mcitedefaultmidpunct}
{\mcitedefaultendpunct}{\mcitedefaultseppunct}\relax
\EndOfBibitem
\bibitem[Ando and Shiraishi(2017)Ando, and Shiraishi]{ando2017spin}
Ando,~Y.; Shiraishi,~M. Spin to charge interconversion phenomena in the
  interface and surface states. \emph{Journal of the Physical Society of Japan}
  \textbf{2017}, \emph{86}, 011001\relax
\mciteBstWouldAddEndPuncttrue
\mciteSetBstMidEndSepPunct{\mcitedefaultmidpunct}
{\mcitedefaultendpunct}{\mcitedefaultseppunct}\relax
\EndOfBibitem
\bibitem[Soumyanarayanan \latin{et~al.}(2016)Soumyanarayanan, Reyren, Fert, and
  Panagopoulos]{soumyanarayanan2016emergent}
Soumyanarayanan,~A.; Reyren,~N.; Fert,~A.; Panagopoulos,~C. Emergent phenomena
  induced by spin--orbit coupling at surfaces and interfaces. \emph{Nature}
  \textbf{2016}, \emph{539}, 509--517\relax
\mciteBstWouldAddEndPuncttrue
\mciteSetBstMidEndSepPunct{\mcitedefaultmidpunct}
{\mcitedefaultendpunct}{\mcitedefaultseppunct}\relax
\EndOfBibitem
\bibitem[Miron \latin{et~al.}(2011)Miron, Garello, Gaudin, Zermatten, Costache,
  Auffret, Bandiera, Rodmacq, Schuhl, and Gambardella]{miron2011perpendicular}
Miron,~I.~M.; Garello,~K.; Gaudin,~G.; Zermatten,~P.-J.; Costache,~M.~V.;
  Auffret,~S.; Bandiera,~S.; Rodmacq,~B.; Schuhl,~A.; Gambardella,~P.
  Perpendicular switching of a single ferromagnetic layer induced by in-plane
  current injection. \emph{Nature} \textbf{2011}, \emph{476}, 189--193\relax
\mciteBstWouldAddEndPuncttrue
\mciteSetBstMidEndSepPunct{\mcitedefaultmidpunct}
{\mcitedefaultendpunct}{\mcitedefaultseppunct}\relax
\EndOfBibitem
\bibitem[Liu \latin{et~al.}(2012)Liu, Pai, Li, Tseng, Ralph, and
  Buhrman]{liu2012spin}
Liu,~L.; Pai,~C.-F.; Li,~Y.; Tseng,~H.; Ralph,~D.; Buhrman,~R. Spin-torque
  switching with the giant spin Hall effect of tantalum. \emph{Science}
  \textbf{2012}, \emph{336}, 555--558\relax
\mciteBstWouldAddEndPuncttrue
\mciteSetBstMidEndSepPunct{\mcitedefaultmidpunct}
{\mcitedefaultendpunct}{\mcitedefaultseppunct}\relax
\EndOfBibitem
\bibitem[Safeer \latin{et~al.}(2016)Safeer, Ju{\'e}, Lopez, Buda-Prejbeanu,
  Auffret, Pizzini, Boulle, Miron, and Gaudin]{safeer2016spin}
Safeer,~C.; Ju{\'e},~E.; Lopez,~A.; Buda-Prejbeanu,~L.; Auffret,~S.;
  Pizzini,~S.; Boulle,~O.; Miron,~I.~M.; Gaudin,~G. Spin--orbit torque
  magnetization switching controlled by geometry. \emph{Nature nanotechnology}
  \textbf{2016}, \emph{11}, 143\relax
\mciteBstWouldAddEndPuncttrue
\mciteSetBstMidEndSepPunct{\mcitedefaultmidpunct}
{\mcitedefaultendpunct}{\mcitedefaultseppunct}\relax
\EndOfBibitem
\bibitem[Pham \latin{et~al.}(2020)Pham, Groen, Manipatruni, Choi, Nikonov,
  Sagasta, Lin, Gosavi, Marty, Hueso, Young, and Casanova]{pham2020spin}
Pham,~V.~T.; Groen,~I.; Manipatruni,~S.; Choi,~W.~Y.; Nikonov,~D.~E.;
  Sagasta,~E.; Lin,~C.-C.; Gosavi,~T.~A.; Marty,~A.; Hueso,~L.~E.;
  Young,~I.~A.; Casanova,~F. Spin–orbit magnetic state readout in scaled
  ferromagnetic/heavy metal nanostructures. \emph{Nature Electronics}
  \textbf{2020}, \emph{3}, 309--315\relax
\mciteBstWouldAddEndPuncttrue
\mciteSetBstMidEndSepPunct{\mcitedefaultmidpunct}
{\mcitedefaultendpunct}{\mcitedefaultseppunct}\relax
\EndOfBibitem
\bibitem[Manipatruni \latin{et~al.}(2019)Manipatruni, Nikonov, Lin, Gosavi,
  Liu, Prasad, Huang, Bonturim, Ramesh, and Young]{manipatruni2019scalable}
Manipatruni,~S.; Nikonov,~D.~E.; Lin,~C.-C.; Gosavi,~T.~A.; Liu,~H.;
  Prasad,~B.; Huang,~Y.-L.; Bonturim,~E.; Ramesh,~R.; Young,~I.~A. Scalable
  energy-efficient magnetoelectric spin--orbit logic. \emph{Nature}
  \textbf{2019}, \emph{565}, 35--42\relax
\mciteBstWouldAddEndPuncttrue
\mciteSetBstMidEndSepPunct{\mcitedefaultmidpunct}
{\mcitedefaultendpunct}{\mcitedefaultseppunct}\relax
\EndOfBibitem
\bibitem[Karube \latin{et~al.}(2016)Karube, Kondou, and
  Otani]{karube2016experimental}
Karube,~S.; Kondou,~K.; Otani,~Y. Experimental observation of spin-to-charge
  current conversion at non-magnetic metal/Bi2O3interfaces. \emph{Applied
  Physics Express} \textbf{2016}, \emph{9}, 033001\relax
\mciteBstWouldAddEndPuncttrue
\mciteSetBstMidEndSepPunct{\mcitedefaultmidpunct}
{\mcitedefaultendpunct}{\mcitedefaultseppunct}\relax
\EndOfBibitem
\bibitem[Kim \latin{et~al.}(2017)Kim, Chen, Karube, Takahashi, Kondou, Tatara,
  and Otani]{kim2017evaluation}
Kim,~J.; Chen,~Y.-T.; Karube,~S.; Takahashi,~S.; Kondou,~K.; Tatara,~G.;
  Otani,~Y. Evaluation of bulk-interface contributions to Edelstein
  magnetoresistance at metal/oxide interfaces. \emph{Phys. Rev. B}
  \textbf{2017}, \emph{96}, 140409\relax
\mciteBstWouldAddEndPuncttrue
\mciteSetBstMidEndSepPunct{\mcitedefaultmidpunct}
{\mcitedefaultendpunct}{\mcitedefaultseppunct}\relax
\EndOfBibitem
\bibitem[Nakayama \latin{et~al.}(2016)Nakayama, Kanno, An, Tashiro, Haku,
  Nomura, and Ando]{nakayama2016rashba}
Nakayama,~H.; Kanno,~Y.; An,~H.; Tashiro,~T.; Haku,~S.; Nomura,~A.; Ando,~K.
  Rashba-Edelstein Magnetoresistance in Metallic Heterostructures. \emph{Phys.
  Rev. Lett.} \textbf{2016}, \emph{117}, 116602\relax
\mciteBstWouldAddEndPuncttrue
\mciteSetBstMidEndSepPunct{\mcitedefaultmidpunct}
{\mcitedefaultendpunct}{\mcitedefaultseppunct}\relax
\EndOfBibitem
\bibitem[Tsai \latin{et~al.}(2018)Tsai, Karube, Kondou, Yamaguchi, Ishii, and
  Otani]{tsai2018clear}
Tsai,~H.; Karube,~S.; Kondou,~K.; Yamaguchi,~N.; Ishii,~F.; Otani,~Y. Clear
  variation of spin splitting by changing electron distribution at non-magnetic
  metal/Bi 2 O 3 interfaces. \emph{Scientific reports} \textbf{2018}, \emph{8},
  1--8\relax
\mciteBstWouldAddEndPuncttrue
\mciteSetBstMidEndSepPunct{\mcitedefaultmidpunct}
{\mcitedefaultendpunct}{\mcitedefaultseppunct}\relax
\EndOfBibitem
\bibitem[Rojas-S\'anchez \latin{et~al.}(2016)Rojas-S\'anchez, Oyarz\'un, Fu,
  Marty, Vergnaud, Gambarelli, Vila, Jamet, Ohtsubo, Taleb-Ibrahimi,
  Le~F\`evre, Bertran, Reyren, George, and Fert]{rojas2016spin}
Rojas-S\'anchez,~J.-C.; Oyarz\'un,~S.; Fu,~Y.; Marty,~A.; Vergnaud,~C.;
  Gambarelli,~S.; Vila,~L.; Jamet,~M.; Ohtsubo,~Y.; Taleb-Ibrahimi,~A.;
  Le~F\`evre,~P.; Bertran,~F.; Reyren,~N.; George,~J.-M.; Fert,~A. Spin to
  Charge Conversion at Room Temperature by Spin Pumping into a New Type of
  Topological Insulator: $\ensuremath{\alpha}$-Sn Films. \emph{Phys. Rev.
  Lett.} \textbf{2016}, \emph{116}, 096602\relax
\mciteBstWouldAddEndPuncttrue
\mciteSetBstMidEndSepPunct{\mcitedefaultmidpunct}
{\mcitedefaultendpunct}{\mcitedefaultseppunct}\relax
\EndOfBibitem
\bibitem[Kondou \latin{et~al.}(2016)Kondou, Yoshimi, Tsukazaki, Fukuma,
  Matsuno, Takahashi, Kawasaki, Tokura, and Otani]{kondou2016fermi}
Kondou,~K.; Yoshimi,~R.; Tsukazaki,~A.; Fukuma,~Y.; Matsuno,~J.; Takahashi,~K.;
  Kawasaki,~M.; Tokura,~Y.; Otani,~Y. Fermi-level-dependent charge-to-spin
  current conversion by Dirac surface states of topological insulators.
  \emph{Nature Physics} \textbf{2016}, \emph{12}, 1027--1031\relax
\mciteBstWouldAddEndPuncttrue
\mciteSetBstMidEndSepPunct{\mcitedefaultmidpunct}
{\mcitedefaultendpunct}{\mcitedefaultseppunct}\relax
\EndOfBibitem
\bibitem[Lesne \latin{et~al.}(2016)Lesne, Fu, Oyarzun, Rojas-S{\'a}nchez, Vaz,
  Naganuma, Sicoli, Attan{\'e}, Jamet, Jacquet, \latin{et~al.}
  others]{lesne2016highly}
Lesne,~E.; Fu,~Y.; Oyarzun,~S.; Rojas-S{\'a}nchez,~J.; Vaz,~D.; Naganuma,~H.;
  Sicoli,~G.; Attan{\'e},~J.-P.; Jamet,~M.; Jacquet,~E., \latin{et~al.}  Highly
  efficient and tunable spin-to-charge conversion through Rashba coupling at
  oxide interfaces. \emph{Nature materials} \textbf{2016}, \emph{15},
  1261--1266\relax
\mciteBstWouldAddEndPuncttrue
\mciteSetBstMidEndSepPunct{\mcitedefaultmidpunct}
{\mcitedefaultendpunct}{\mcitedefaultseppunct}\relax
\EndOfBibitem
\bibitem[Vaz \latin{et~al.}(2019)Vaz, No{\"e}l, Johansson, G{\"o}bel, Bruno,
  Singh, Mckeown-Walker, Trier, Vicente-Arche, Sander, \latin{et~al.}
  others]{vaz2019mapping}
Vaz,~D.~C.; No{\"e}l,~P.; Johansson,~A.; G{\"o}bel,~B.; Bruno,~F.~Y.;
  Singh,~G.; Mckeown-Walker,~S.; Trier,~F.; Vicente-Arche,~L.~M.; Sander,~A.,
  \latin{et~al.}  Mapping spin--charge conversion to the band structure in a
  topological oxide two-dimensional electron gas. \emph{Nature materials}
  \textbf{2019}, \emph{18}, 1187--1193\relax
\mciteBstWouldAddEndPuncttrue
\mciteSetBstMidEndSepPunct{\mcitedefaultmidpunct}
{\mcitedefaultendpunct}{\mcitedefaultseppunct}\relax
\EndOfBibitem
\bibitem[Linder and Yokoyama(2011)Linder, and Yokoyama]{LinYok2011}
Linder,~J.; Yokoyama,~T. Spin Current in Generic Hybrid Structures due to
  Interfacial Spin-Orbit Scattering. \emph{Phys. Rev. Lett.} \textbf{2011},
  \emph{106}, 237201\relax
\mciteBstWouldAddEndPuncttrue
\mciteSetBstMidEndSepPunct{\mcitedefaultmidpunct}
{\mcitedefaultendpunct}{\mcitedefaultseppunct}\relax
\EndOfBibitem
\bibitem[Tokatly \latin{et~al.}(2015)Tokatly, Krasovskii, and
  Vignale]{TokKraVig2015PRB}
Tokatly,~I.~V.; Krasovskii,~E.~E.; Vignale,~G. Current-induced spin
  polarization at the surface of metallic films: A theorem and an \textit{ab
  initio} calculation. \emph{Phys. Rev. B} \textbf{2015}, \emph{91},
  035403\relax
\mciteBstWouldAddEndPuncttrue
\mciteSetBstMidEndSepPunct{\mcitedefaultmidpunct}
{\mcitedefaultendpunct}{\mcitedefaultseppunct}\relax
\EndOfBibitem
\bibitem[Borge and Tokatly(2017)Borge, and Tokatly]{BorTok2017PRB}
Borge,~J.; Tokatly,~I.~V. Ballistic spin transport in the presence of
  interfaces with strong spin-orbit coupling. \emph{Phys. Rev. B}
  \textbf{2017}, \emph{96}, 115445\relax
\mciteBstWouldAddEndPuncttrue
\mciteSetBstMidEndSepPunct{\mcitedefaultmidpunct}
{\mcitedefaultendpunct}{\mcitedefaultseppunct}\relax
\EndOfBibitem
\bibitem[Borge and Tokatly(2019)Borge, and Tokatly]{borge2019boundary}
Borge,~J.; Tokatly,~I.~V. Boundary conditions for spin and charge diffusion in
  the presence of interfacial spin-orbit coupling. \emph{Phys. Rev. B}
  \textbf{2019}, \emph{99}, 241401\relax
\mciteBstWouldAddEndPuncttrue
\mciteSetBstMidEndSepPunct{\mcitedefaultmidpunct}
{\mcitedefaultendpunct}{\mcitedefaultseppunct}\relax
\EndOfBibitem
\bibitem[Sanz-Fern\'andez \latin{et~al.}(2019)Sanz-Fern\'andez, Borge, Tokatly,
  and Bergeret]{sanz2019nonlocal}
Sanz-Fern\'andez,~C.; Borge,~J.; Tokatly,~I.~V.; Bergeret,~F.~S. Nonlocal
  magnetolectric effects in diffusive conductors with spatially inhomogeneous
  spin-orbit coupling. \emph{Phys. Rev. B} \textbf{2019}, \emph{100},
  195406\relax
\mciteBstWouldAddEndPuncttrue
\mciteSetBstMidEndSepPunct{\mcitedefaultmidpunct}
{\mcitedefaultendpunct}{\mcitedefaultseppunct}\relax
\EndOfBibitem
\bibitem[Amin \latin{et~al.}(2018)Amin, Zemen, and Stiles]{amin2018interface}
Amin,~V.~P.; Zemen,~J.; Stiles,~M.~D. Interface-Generated Spin Currents.
  \emph{Phys. Rev. Lett.} \textbf{2018}, \emph{121}, 136805\relax
\mciteBstWouldAddEndPuncttrue
\mciteSetBstMidEndSepPunct{\mcitedefaultmidpunct}
{\mcitedefaultendpunct}{\mcitedefaultseppunct}\relax
\EndOfBibitem
\bibitem[Amin and Stiles(2016)Amin, and Stiles]{amin2016phenomenology}
Amin,~V.~P.; Stiles,~M.~D. Spin transport at interfaces with spin-orbit
  coupling: Phenomenology. \emph{Phys. Rev. B} \textbf{2016}, \emph{94},
  104420\relax
\mciteBstWouldAddEndPuncttrue
\mciteSetBstMidEndSepPunct{\mcitedefaultmidpunct}
{\mcitedefaultendpunct}{\mcitedefaultseppunct}\relax
\EndOfBibitem
\bibitem[Amin and Stiles(2016)Amin, and Stiles]{amin2016formalism}
Amin,~V.~P.; Stiles,~M.~D. Spin transport at interfaces with spin-orbit
  coupling: Formalism. \emph{Phys. Rev. B} \textbf{2016}, \emph{94},
  104419\relax
\mciteBstWouldAddEndPuncttrue
\mciteSetBstMidEndSepPunct{\mcitedefaultmidpunct}
{\mcitedefaultendpunct}{\mcitedefaultseppunct}\relax
\EndOfBibitem
\bibitem[Takahashi and Maekawa(2003)Takahashi, and Maekawa]{takahashi2003spin}
Takahashi,~S.; Maekawa,~S. Spin injection and detection in magnetic
  nanostructures. \emph{Phys. Rev. B} \textbf{2003}, \emph{67}, 052409\relax
\mciteBstWouldAddEndPuncttrue
\mciteSetBstMidEndSepPunct{\mcitedefaultmidpunct}
{\mcitedefaultendpunct}{\mcitedefaultseppunct}\relax
\EndOfBibitem
\bibitem[Niimi \latin{et~al.}(2014)Niimi, Suzuki, Kawanishi, Omori, Valet,
  Fert, and Otani]{niimi2014extrinsic}
Niimi,~Y.; Suzuki,~H.; Kawanishi,~Y.; Omori,~Y.; Valet,~T.; Fert,~A.; Otani,~Y.
  Extrinsic spin Hall effects measured with lateral spin valve structures.
  \emph{Phys. Rev. B} \textbf{2014}, \emph{89}, 054401\relax
\mciteBstWouldAddEndPuncttrue
\mciteSetBstMidEndSepPunct{\mcitedefaultmidpunct}
{\mcitedefaultendpunct}{\mcitedefaultseppunct}\relax
\EndOfBibitem
\bibitem[Isshiki \latin{et~al.}(2019)Isshiki, Muduli, Kim, Kondou, and
  Otani]{isshiki2019experimentally}
Isshiki,~H.; Muduli,~P.; Kim,~J.; Kondou,~K.; Otani,~Y. Experimentally
  determined correlation between direct and inverse Edelstein effects at
  Bi2O3/Cu interface by means of spin absorption method using lateral spin
  valve structure. \emph{arXiv preprint arXiv:1901.03095} \textbf{2019}, \relax
\mciteBstWouldAddEndPunctfalse
\mciteSetBstMidEndSepPunct{\mcitedefaultmidpunct}
{}{\mcitedefaultseppunct}\relax
\EndOfBibitem
\bibitem[Pham \latin{et~al.}(2016)Pham, Vila, Zahnd, Marty, Savero-Torres,
  Jamet, and Attané]{pham2016ferro}
Pham,~V.~T.; Vila,~L.; Zahnd,~G.; Marty,~A.; Savero-Torres,~W.; Jamet,~M.;
  Attané,~J.-P. Ferromagnetic/Nonmagnetic Nanostructures for the Electrical
  Measurement of the Spin Hall Effect. \emph{Nano Letters} \textbf{2016},
  \emph{16}, 6755--6760, PMID: 27712075\relax
\mciteBstWouldAddEndPuncttrue
\mciteSetBstMidEndSepPunct{\mcitedefaultmidpunct}
{\mcitedefaultendpunct}{\mcitedefaultseppunct}\relax
\EndOfBibitem
\bibitem[Kimura \latin{et~al.}(2005)Kimura, Hamrle, and
  Otani]{kimura2005estimation}
Kimura,~T.; Hamrle,~J.; Otani,~Y. Estimation of spin-diffusion length from the
  magnitude of spin-current absorption: Multiterminal
  ferromagnetic/nonferromagnetic hybrid structures. \emph{Phys. Rev. B}
  \textbf{2005}, \emph{72}, 014461\relax
\mciteBstWouldAddEndPuncttrue
\mciteSetBstMidEndSepPunct{\mcitedefaultmidpunct}
{\mcitedefaultendpunct}{\mcitedefaultseppunct}\relax
\EndOfBibitem
\bibitem[Kimura \latin{et~al.}(2004)Kimura, Hamrle, Otani, Tsukagoshi, and
  Aoyagi]{kimura2004spin}
Kimura,~T.; Hamrle,~J.; Otani,~Y.; Tsukagoshi,~K.; Aoyagi,~Y. Spin-dependent
  boundary resistance in the lateral spin-valve structure. \emph{Applied
  Physics Letters} \textbf{2004}, \emph{85}, 3501--3503\relax
\mciteBstWouldAddEndPuncttrue
\mciteSetBstMidEndSepPunct{\mcitedefaultmidpunct}
{\mcitedefaultendpunct}{\mcitedefaultseppunct}\relax
\EndOfBibitem
\bibitem[Isasa \latin{et~al.}(2016)Isasa, Mart\'{\i}nez-Velarte, Villamor,
  Mag\'en, Morell\'on, De~Teresa, Ibarra, Vignale, Chulkov, Krasovskii, Hueso,
  and Casanova]{isasa2016origin}
Isasa,~M.; Mart\'{\i}nez-Velarte,~M.~C.; Villamor,~E.; Mag\'en,~C.;
  Morell\'on,~L.; De~Teresa,~J.~M.; Ibarra,~M.~R.; Vignale,~G.; Chulkov,~E.~V.;
  Krasovskii,~E.~E.; Hueso,~L.~E.; Casanova,~F. Origin of inverse
  Rashba-Edelstein effect detected at the Cu/Bi interface using lateral spin
  valves. \emph{Phys. Rev. B} \textbf{2016}, \emph{93}, 014420\relax
\mciteBstWouldAddEndPuncttrue
\mciteSetBstMidEndSepPunct{\mcitedefaultmidpunct}
{\mcitedefaultendpunct}{\mcitedefaultseppunct}\relax
\EndOfBibitem
\bibitem[Sagasta \latin{et~al.}(2017)Sagasta, Omori, Isasa, Otani, Hueso, and
  Casanova]{sagasta2017spin}
Sagasta,~E.; Omori,~Y.; Isasa,~M.; Otani,~Y.; Hueso,~L.~E.; Casanova,~F. Spin
  diffusion length of Permalloy using spin absorption in lateral spin valves.
  \emph{Applied Physics Letters} \textbf{2017}, \emph{111}, 082407\relax
\mciteBstWouldAddEndPuncttrue
\mciteSetBstMidEndSepPunct{\mcitedefaultmidpunct}
{\mcitedefaultendpunct}{\mcitedefaultseppunct}\relax
\EndOfBibitem
\bibitem[Tsai \latin{et~al.}(2019)Tsai, Kondou, and Otani]{tsai2019enhanced}
Tsai,~H.; Kondou,~K.; Otani,~Y. Enhanced spin-to-charge current conversion at
  metal/oxide interfaces by lowering the temperature. \emph{Japanese Journal of
  Applied Physics} \textbf{2019}, \emph{58}, 110907\relax
\mciteBstWouldAddEndPuncttrue
\mciteSetBstMidEndSepPunct{\mcitedefaultmidpunct}
{\mcitedefaultendpunct}{\mcitedefaultseppunct}\relax
\EndOfBibitem
\bibitem[S{\'a}nchez \latin{et~al.}(2013)S{\'a}nchez, Vila, Desfonds,
  Gambarelli, Attan{\'e}, De~Teresa, Mag{\'e}n, and Fert]{sanchez2013spin}
S{\'a}nchez,~J.~R.; Vila,~L.; Desfonds,~G.; Gambarelli,~S.; Attan{\'e},~J.;
  De~Teresa,~J.; Mag{\'e}n,~C.; Fert,~A. Spin-to-charge conversion using Rashba
  coupling at the interface between non-magnetic materials. \emph{Nature
  communications} \textbf{2013}, \emph{4}, 1--7\relax
\mciteBstWouldAddEndPuncttrue
\mciteSetBstMidEndSepPunct{\mcitedefaultmidpunct}
{\mcitedefaultendpunct}{\mcitedefaultseppunct}\relax
\EndOfBibitem
\bibitem[Chen \latin{et~al.}(2013)Chen, Takahashi, Nakayama, Althammer,
  Goennenwein, Saitoh, and Bauer]{chen2013theory}
Chen,~Y.-T.; Takahashi,~S.; Nakayama,~H.; Althammer,~M.; Goennenwein,~S. T.~B.;
  Saitoh,~E.; Bauer,~G. E.~W. Theory of spin Hall magnetoresistance.
  \emph{Phys. Rev. B} \textbf{2013}, \emph{87}, 144411\relax
\mciteBstWouldAddEndPuncttrue
\mciteSetBstMidEndSepPunct{\mcitedefaultmidpunct}
{\mcitedefaultendpunct}{\mcitedefaultseppunct}\relax
\EndOfBibitem
\bibitem[Nakayama \latin{et~al.}(2013)Nakayama, Althammer, Chen, Uchida,
  Kajiwara, Kikuchi, Ohtani, Gepr{\"a}gs, Opel, Takahashi, \latin{et~al.}
  others]{nakayama2013spin}
Nakayama,~H.; Althammer,~M.; Chen,~Y.-T.; Uchida,~K.; Kajiwara,~Y.;
  Kikuchi,~D.; Ohtani,~T.; Gepr{\"a}gs,~S.; Opel,~M.; Takahashi,~S.,
  \latin{et~al.}  Spin Hall magnetoresistance induced by a nonequilibrium
  proximity effect. \emph{Physical review letters} \textbf{2013}, \emph{110},
  206601\relax
\mciteBstWouldAddEndPuncttrue
\mciteSetBstMidEndSepPunct{\mcitedefaultmidpunct}
{\mcitedefaultendpunct}{\mcitedefaultseppunct}\relax
\EndOfBibitem
\bibitem[Zhang \latin{et~al.}(2019)Zhang, Bergeret, and
  Golovach]{zhang2019theory}
Zhang,~X.-P.; Bergeret,~F.~S.; Golovach,~V.~N. Theory of Spin Hall
  Magnetoresistance from a Microscopic Perspective. \emph{Nano Letters}
  \textbf{2019}, \emph{19}, 6330--6337, PMID: 31378061\relax
\mciteBstWouldAddEndPuncttrue
\mciteSetBstMidEndSepPunct{\mcitedefaultmidpunct}
{\mcitedefaultendpunct}{\mcitedefaultseppunct}\relax
\EndOfBibitem
\bibitem[Brataas \latin{et~al.}(2000)Brataas, Nazarov, and
  Bauer]{brataas2000finite}
Brataas,~A.; Nazarov,~Y.~V.; Bauer,~G.~E. Finite-element theory of transport in
  ferromagnet--normal metal systems. \emph{Physical Review Letters}
  \textbf{2000}, \emph{84}, 2481\relax
\mciteBstWouldAddEndPuncttrue
\mciteSetBstMidEndSepPunct{\mcitedefaultmidpunct}
{\mcitedefaultendpunct}{\mcitedefaultseppunct}\relax
\EndOfBibitem
\bibitem[Vlietstra \latin{et~al.}(2013)Vlietstra, Shan, Castel, Ben~Youssef,
  Bauer, and van Wees]{vlietstra2013exchange}
Vlietstra,~N.; Shan,~J.; Castel,~V.; Ben~Youssef,~J.; Bauer,~G. E.~W.; van
  Wees,~B.~J. Exchange magnetic field torques in YIG/Pt bilayers observed by
  the spin-Hall magnetoresistance. \emph{Applied Physics Letters}
  \textbf{2013}, \emph{103}, 032401\relax
\mciteBstWouldAddEndPuncttrue
\mciteSetBstMidEndSepPunct{\mcitedefaultmidpunct}
{\mcitedefaultendpunct}{\mcitedefaultseppunct}\relax
\EndOfBibitem
\bibitem[Althammer \latin{et~al.}(2013)Althammer, Meyer, Nakayama, Schreier,
  Altmannshofer, Weiler, Huebl, Gepr\"ags, Opel, Gross, Meier, Klewe, Kuschel,
  Schmalhorst, Reiss, Shen, Gupta, Chen, Bauer, Saitoh, and
  Goennenwein]{althammer2013quantitative}
Althammer,~M. \latin{et~al.}  Quantitative study of the spin Hall
  magnetoresistance in ferromagnetic insulator/normal metal hybrids.
  \emph{Phys. Rev. B} \textbf{2013}, \emph{87}, 224401\relax
\mciteBstWouldAddEndPuncttrue
\mciteSetBstMidEndSepPunct{\mcitedefaultmidpunct}
{\mcitedefaultendpunct}{\mcitedefaultseppunct}\relax
\EndOfBibitem
\bibitem[Kosub \latin{et~al.}(2018)Kosub, Vélez, Gomez-Perez, Hueso,
  Fassbender, Casanova, and Makarov]{kosub2018anomalous}
Kosub,~T.; Vélez,~S.; Gomez-Perez,~J.~M.; Hueso,~L.~E.; Fassbender,~J.;
  Casanova,~F.; Makarov,~D. Anomalous Hall-like transverse magnetoresistance in
  Au thin films on Y3Fe5O12. \emph{Applied Physics Letters} \textbf{2018},
  \emph{113}, 222409\relax
\mciteBstWouldAddEndPuncttrue
\mciteSetBstMidEndSepPunct{\mcitedefaultmidpunct}
{\mcitedefaultendpunct}{\mcitedefaultseppunct}\relax
\EndOfBibitem
\bibitem[Das \latin{et~al.}(2019)Das, Dejene, van Wees, and
  Vera-Marun]{das2019temperature}
Das,~K.~S.; Dejene,~F.~K.; van Wees,~B.~J.; Vera-Marun,~I.~J. Temperature
  dependence of the effective spin-mixing conductance probed with lateral
  non-local spin valves. \emph{Applied Physics Letters} \textbf{2019},
  \emph{114}, 072405\relax
\mciteBstWouldAddEndPuncttrue
\mciteSetBstMidEndSepPunct{\mcitedefaultmidpunct}
{\mcitedefaultendpunct}{\mcitedefaultseppunct}\relax
\EndOfBibitem
\bibitem[Villamor \latin{et~al.}(2015)Villamor, Isasa, V\'elez, Bedoya-Pinto,
  Vavassori, Hueso, Bergeret, and Casanova]{villamor2015modulation}
Villamor,~E.; Isasa,~M.; V\'elez,~S.; Bedoya-Pinto,~A.; Vavassori,~P.;
  Hueso,~L.~E.; Bergeret,~F.~S.; Casanova,~F. Modulation of pure spin currents
  with a ferromagnetic insulator. \emph{Phys. Rev. B} \textbf{2015}, \emph{91},
  020403\relax
\mciteBstWouldAddEndPuncttrue
\mciteSetBstMidEndSepPunct{\mcitedefaultmidpunct}
{\mcitedefaultendpunct}{\mcitedefaultseppunct}\relax
\EndOfBibitem
\bibitem[Cornelissen \latin{et~al.}(2016)Cornelissen, Peters, Bauer, Duine, and
  van Wees]{cornelissen2016magnon}
Cornelissen,~L.~J.; Peters,~K. J.~H.; Bauer,~G. E.~W.; Duine,~R.~A.; van
  Wees,~B.~J. Magnon spin transport driven by the magnon chemical potential in
  a magnetic insulator. \emph{Phys. Rev. B} \textbf{2016}, \emph{94},
  014412\relax
\mciteBstWouldAddEndPuncttrue
\mciteSetBstMidEndSepPunct{\mcitedefaultmidpunct}
{\mcitedefaultendpunct}{\mcitedefaultseppunct}\relax
\EndOfBibitem
\bibitem[Vlietstra \latin{et~al.}(2013)Vlietstra, Shan, Castel, van Wees, and
  Ben~Youssef]{vlietstra2013spin}
Vlietstra,~N.; Shan,~J.; Castel,~V.; van Wees,~B.~J.; Ben~Youssef,~J. Spin-Hall
  magnetoresistance in platinum on yttrium iron garnet: Dependence on platinum
  thickness and in-plane/out-of-plane magnetization. \emph{Phys. Rev. B}
  \textbf{2013}, \emph{87}, 184421\relax
\mciteBstWouldAddEndPuncttrue
\mciteSetBstMidEndSepPunct{\mcitedefaultmidpunct}
{\mcitedefaultendpunct}{\mcitedefaultseppunct}\relax
\EndOfBibitem
\end{mcitethebibliography}

%FIGURES


\end{document}